\documentclass[prc,aps,twocolumn,superscriptaddress,nofootinbib,showpacs,showkeys]{revtex4} 
\usepackage{graphicx,epsf,amsfonts,amssymb,amsbsy,pstricks} 
 
\newcommand{\be} {\begin{eqnarray*}} 
\newcommand{\ee} {\end{eqnarray*}} 
 
\newcommand{\bcen}{\begin{center}} 
\newcommand{\ecen}{\end{center}} 
\newcommand{\beq}{\begin{equation}} 
\newcommand{\eeq}{\end{equation}} 
\newcommand{\bea}{\begin{eqnarray}} 
\newcommand{\eea}{\end{eqnarray}} 
\newcommand{\ba}{\begin{array}} 
\newcommand{\ea}{\end{array}} 
\newcommand{\bann}{\begin{eqnarray*}} 
\newcommand{\eann}{\end{eqnarray*}}

\newcommand{\lsim}[1]{ 
\setlength{\unitlength}{12pt} 
\begin{picture}(1.4,1.) 
\put(.7,-0.3){\makebox(0.0,1.)[t]{$<$}} 
\put(.7,-0.3){\makebox(0.0,1.)[b]{$\sim$}} 
\end{picture}#1} 
 
\begin{document} 
\title{Nonlocality effects on Color Spin Locking condensates} 
 
\author{D. N. Aguilera} 
\email{deborah.aguilera@ua.es} 
\affiliation{ 
Department of Applied Physics, University of Alicante, Ap. Correus 99, 
03080 Alicante, Spain} 
 
\author{D. Blaschke} 
\email{david.blaschke@uni-rostock.de} 
\affiliation{Institute for Theoretical Physics, University of Wroclaw, 
Max Born pl. 3, 50-204 Wroclaw, Poland} 
\affiliation{Institut f\"ur Physik, Universit\"at Rostock, 
Universit\"atsplatz 3, 18051 Rostock, Germany} 
\affiliation{Bogoliubov  Laboratory of Theoretical Physics, JINR Dubna, 
Joliot-Curie Street 6, 141980  Dubna, Russia} 
 
\author{H. Grigorian} 
\email{hovik.grigorian@uni-rostock.de} 
\affiliation{Laboratory for Information Technologies, 
JINR Dubna, Joliot-Burie Street 6, 141980 Dubna, Russia} 
\affiliation{Institut f\"ur Physik, Universit\"at Rostock, 
Universit\"atsplatz 3, 18051 Rostock, Germany} 
\affiliation{Department of Physics, Yerevan State University, Alex Manookian 
str. 1, 375047 Yerevan, Armenia} 
 
\author{N. N. Scoccola} 
\email{scoccola@tandar.cnea.gov.ar} 
\affiliation{Physics Department, Comisi\'on Nacional de 
Energ\'{\i}a At\'omica, Av. Libertador 8250, 1429 Buenos Aires, Argentina} 
\affiliation{Universidad Favaloro, Sol{\'\i}s 453, 1078 Buenos Aires, 
Argentina} 
\affiliation{CONICET, Rivadavia 1917, 1033 Buenos Aires, Argentina} 
\date{\today} 
\begin{abstract} 
We consider the color spin locking (CSL) phase of two-flavor quark 
matter at zero temperature for nonlocal instantaneous, separable 
interactions. 
We employ a Lorentzian-type form factor allowing a parametric 
interpolation between the sharp (Nambu-Jona-Lasinio (NJL) model) 
and very smooth (e.g. Gaussian) cut-off models for systematic studies of 
the influence 
on the CSL condensate the deviation from the NJL model entails. 
This smoothing of the NJL model form factor shows advantageous features 
for the phenomenology of compact stars: (i) a lowering of the 
critical chemical potential for the onset of the chiral phase 
transition as a prerequisite for stability of hybrid stars with 
extended quark matter cores and (ii) a reduction of the smallest 
pairing gap to the order of 100 keV, being in the range of values
interesting for phenomenological studies of hybrid star cooling 
evolution. 
 
\pacs{04.40.Dg, 12.38.Mh, 26.60.+c, 97.60.Jd} 
 
\keywords{ 
spin-one condensates, CSL phase, chiral quark model, nonlocal interactions, 
compact star cooling} 
\end{abstract} 
\maketitle 
\section{Introduction} 
 
Recently, the investigation of color superconducting phases 
in cold dense quark matter 
has received much attention \cite{Buballa:2003qv,Alford:2002wf,Casalbuoni:2003wh,Schafer:2003vz,Rapp:1999qa,Rajagopal:2000wf}, 
in particular due to the possible consequences for the physics 
of compact stars \cite{Blaschke:2001uj,Blaschke:2006xt}. 
From the point of view of observational constraints on 
quark matter and color superconductivity in compact stars 
the cooling characteristics play a central role. 
It has been shown that the occurrence of a normal quark matter 
core would lead to a conflict with observations since the 
direct Urca (DU) process in normal quark matter would lead to enhanced cooling 
in disagreement with the data \cite{Blaschke:2000dy,Grigorian:2004jq}. 
The DU conflict would be solved provided no ungapped quark modes 
occur in the quark core. 
{This has been demonstrated on the 
example} of a hypothetical pairing channel {(X-gap)} 
for the quark color which is ungapped in the 2SC phase (2SC+X phase) 
\cite{Grigorian:2004jq,Blaschke:2005dc}. 
However, the microscopic origin of the X-gap could not yet be specified. 
A microscopically well-defined pairing pattern which could solve the quark 
DU problem would be the CSL phase \cite{Aguilera:2005tg} corresponding to a 
spin-one condensate \cite{Schafer:2000tw,Alford:2002rz,Schmitt:2004et}. 
A prerequisite for the realization of this pairing pattern in quark matter 
would be a sufficient flavor asymmetry to prevent the u-d pairing in the 
otherwise dominant scalar diquark channel of the 2SC phase. 
It has been demonstrated that under neutron star conditions the 2SC phase 
is indeed rather fragile and may not be realized for moderate coupling 
strengths \cite{Aguilera:2004ag}. 
Thus the CSL phase becomes particularly interesting  for the solution of the 
quark DU cooling problem, and corresponding simulations will be performed as 
soon as the cooling regulators such as emissivities, specific heat and thermal 
conductivity will be provided. First steps in this direction have been 
made recently \cite{Schmitt:2005wg,Jaikumar:2005hy}. 
 
Most of the calculations of QCD superconducting phases have been 
done using the sharp cut-off NJL model (see Ref.\cite{Buballa:2003qv} 
and references therein). However, lattice QCD calculations 
\cite{Parappilly:2005ei} indicate that quark interactions should act over 
a certain range in the momentum space, and various approaches to include 
nonlocality effects beyond the NJL model have been suggested \cite{Rip97}. 
We refer to nonlocal separable interaction models as introduced, e.g., in the 
works \cite{Schmidt:1994di,Gocke:2001ri,Aguilera:2003yi,Duhau:2004pq,Blaschke:2004cc,GomezDumm:2005hy} 
and references therein, where it has been concluded that smoothing the cutoff 
leads to a reduction of the chiral condensate and a lowering of the 
critical temperature for the chiral phase transition. 
The question arises for the effects of nonlocality on the spin-one gaps, to be 
explored by varying the form factor of the quark interaction from a sharp cutoff 
in the NJL model to smoothly decreasing form such as a Gaussian. 
First exploratory calculations reported in  \cite{Aguilera:2005uf} have shown 
that the nonlocality could lead to a sizeable decrease of the energy gaps. 
 
In this paper, we investigate the robustness of CSL pairing 
against a modification of the {sharp} cut-off (NJL) in a systematic 
way by employing a separable, instantaneous interaction with a 
Lorentzian-type interaction which allows to interpolate between 
the NJL case and very smooth interaction form factors of, e.g., the 
Gaussian type. This investigation is performed on the basis of 
recently developed parameterizations for the instantaneous three 
flavor case \cite{Grigorian:2006qe}. 
 
 
\section{Nonlocal chiral quark model for the color-spin locking (CSL) phase} 
 
We investigate a nonlocal chiral quark model  in which the quark 
interaction is 
represented in a separable way by introducing form factor functions $g(p)$ in 
the bilinears of the current-current interaction terms in the Lagrangian 
\cite{Aguilera:2004ag,Schmidt:1994di,Grigorian:2003vi}. 
It is assumed that this four-fermion interaction is instantaneous 
and therefore the form factors do not depend on the energy but only on the 
modulus of the three momentum $p=|\vec p|$. 
The ansatz for the s-wave, single flavor diquark condensate characterizing the 
CSL phase as introduced in Ref.~\cite{Aguilera:2005tg} 
\footnote{{Note that such ansatz differs from the one in \cite{Schmitt:2004et}, 
where the states are constructed with total angular momentum one.}} 
is a scalar product (locking) of the 
three-vector of antisymmetric color matrices 
$(\lambda_2, \lambda_5, \lambda_7)$ with the three-vector of 
Dirac spin matrices $(\gamma_3,\gamma_2,\gamma_1)$. 
Thus, the corresponding gap matrix $\hat \Delta$ for the CSL phase reads 
\bea 
\hat\Delta= 
\Delta(\gamma_3\lambda_2+\gamma_2\lambda_5+\gamma_1\lambda_7)~. 
\label{hatD_csl} 
\eea 
Since the two flavor channels decouple, 
{the quark thermodynamical potential 
can be decomposed into single-flavor components} 
\bea 
\Omega_q(T,\{\mu_f\}) &=& \sum_{f=u,d} \Omega(T,\mu_f)~, 
\label{omegaq} 
\eea 
{and it is sufficient to consider in the 
following the contribution of a single flavor only}, which in 
the mean field approximation is given by 
\bea 
\Omega(T,\mu)&=& 
 \frac{\phi^2}{8G} 
+3\frac{\Delta^2}{8H_v} 
\nonumber\\ 
&-& T \sum_{n} \int \frac{d^3p}{(2\pi)^3} 
\frac{1}{2}{\rm Tr}\ln\Big({\frac{1}{T}S^{-1}(i\omega_n,\vec p\,)}\Big), 
\label{Omega} 
\eea 
where $\mu$ stands for the chemical potential of that flavor. 
The first two terms are quadratic contributions  of the {mean field 
values $\phi$ and $\Delta$ of the order parameter fields} that signal 
chiral symmetry breaking and CSL superconductivity, respectively. 
Their denominators contain the coupling constants $G$ and $H_v$ in the 
corresponding channel. 
In (\ref{Omega}) the sum is over fermionic Matsubara 
frequencies $\omega_n = (2n+1)\pi T$, and the trace is over Dirac, color and 
Nambu-Gorkov indices. 
 
In our nonlocal extension, the inverse fermion propagator 
differs from the NJL model case by momentum 
dependent form factors $g(p)$ modifying the mesonic and diquark mean fields 
\beq 
S^{-1}(p)= 
\left( 
\begin{array}{cc} 
 \not\!p +\mu\gamma^0-M(p)& 
g(p)\hat\Delta\\ 
-{g(p)\hat\Delta}^\dagger& 
\not\!p -\mu\gamma^0- M(p) 
\end{array} 
\right) 
\label{InvProp} 
\eeq 
where  $M(p)$ is 
the dynamical quark mass function 
\bea 
M(p)=m+g(p)\phi~. 
\label{Mass} 
\eea 
 
Note that although the first two terms in Eq.~(\ref{Omega}) do not have any explicit dependence on the form 
factors, the quantities $\phi$ and $\Delta$ do depend 
implicitly on them through the gap equations,  see Eq.~(\ref{GE_csl}) below. 
 
After evaluation of the trace \cite{Aguilera:2005tg} and Matsubara summation 
the thermodynamical potential takes the form 
\bea 
\Omega(T,\mu) &=& \frac{\phi^2}{8G} + 3\frac{\Delta^2}{8H_v} 
\nonumber\\ 
&-& 
\sum_{k=1}^6 \int \frac{d^3p}{(2\pi)^3} 
\left[E_{k}(p)+2T\ln(1+e^{-E_{k}(p)/T})\right],\nonumber\\ 
\eea 
{where $E_{k}(p)$ denote the excitation energies for the 
modes $k=1\dots 6$. 
{The odd (even) indices denote particle (antiparticle) excitations 
corresponding each to a triplet of spin-one eigenstates.} 
All modes have a gap in the excitation spectrum and can be brought 
into {a} standard form, which for $E_1(p)$ reads } 
\bea 
E_{1}^2(p)&=& (\varepsilon_{{\rm eff}}(p)-\mu_{{\rm eff}}(p))^2+ 
\Delta_{{\rm eff}}^2(p) ~, 
\label{dispersion_1} 
\eea 
with the {effective} quantities 
\bea 
\varepsilon_{{\rm eff}}(p)&=& \sqrt{p\,^2+M_{\rm eff}^2(p)}~,\\ 
M_{\rm eff}(p) &=& \frac{\mu}{\mu_{{\rm eff}}(p)}M(p)~, \\ 
\mu_{{\rm eff}}(p)&=&\mu\sqrt{1+\Delta^2 g^2(p)/\mu^2}~,\\ 
\Delta_{\rm eff}(p)&=& \frac{M(p)}{\mu_{{\rm eff}}(p)} \Delta\, g(p)~, 
\label{effective2} 
\eea 
and for $E_{3,5}(p)$ is given by 
\bea 
\label{E_3_5} 
E_{3,5}^2(p)&=& (\varepsilon(p)-\mu)^2+ 
a_{3,5}(p)\Delta^2 g^2(p)~, 
\eea 
with the momentum-dependent coefficients 
\bea 
a_{3,5}(p)&=&\frac{1}{2}\left[5-\frac{p^2}{\varepsilon(p)\mu} 
\pm \sqrt{\left(1-\frac{p^2}{\varepsilon(p)\mu}\right)^2 
+8\frac{M^2(p)}{\varepsilon^2(p)}}\right],\nonumber\\ 
\label{a_3_5} 
\eea 
where $\varepsilon(p)=\sqrt{p^{2}+M^2(p)}~$. 
The remaining modes $E_{2,4,6}(p)$ are obtained from 
$E_{1,3,5}(p)$ by changing $\mu \rightarrow -\mu$ in 
Eqs. (\ref{dispersion_1})-(\ref{a_3_5}). 
Note that the modes  $E_{1,2}(p)$ correspond to the vanishing 
z-projection of the spin, $S_z=0$, thus being inert against an 
external B-field. 
The remaining modes corresponding to $S_z=\pm 1$ are expected to get 
shifted (Zeeman effect). 
 
For given values of $T$ and $\mu$, the global minimum of 
$\Omega(T,\mu)$ 
in the space of the order parameters $\phi$ and $\Delta$ corresponds to 
the thermodynamical equilibrium state. 
We obtain this state by comparing solutions of the gap equations 
\bea 
\frac{\delta\Omega(T,\mu)}{\delta\phi}= 
\frac{\delta\Omega(T,\mu)}{\delta\Delta}=0 ~. 
\label{GE_csl} 
\eea 
We present results for the case of vanishing temperature and finite chemical 
potential in the next section. 
 
\section{Model calculations} 
 
\subsection{Form factors and their parameters} 
\begin{figure}[bth] 
  \begin{center} 
    \includegraphics[width=0.9\linewidth,height=0.4\textheight,angle = -90]{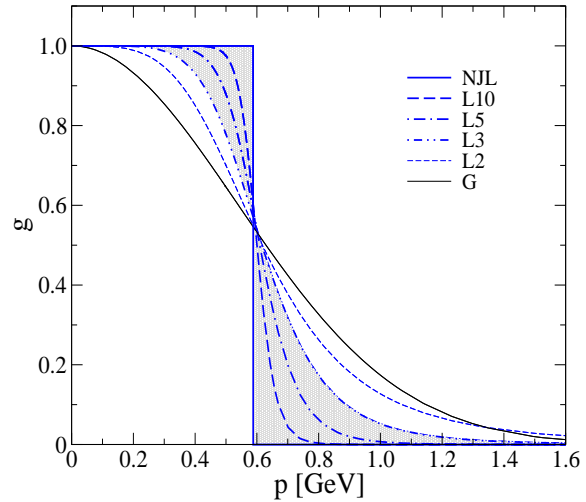} 
    \caption{Form factors used to represent the nonlocality in the momentum 
space. Decreasing $\alpha$ in the Lorentzian-type form factor 
(notation L$\alpha$) causes deviations from the NJL model in a systematic way.} 
    \label{FF_fig} 
  \end{center} 
\end{figure}

In (\ref{InvProp}) and (\ref{Mass}) we have introduced the same 
form factors $g(p)$ to represent the nonlocality of the 
interaction in the meson ($q \bar q$) and diquark ($qq$) channels. 
In our calculations we use the sharp cutoff (NJL), Lorentzian with 
integer parameter $\alpha$ (L$\alpha$) and  Gaussian (G), form 
factors defined as 
\begin{eqnarray} 
\label{NF} 
g_{\rm NJL}(p) &=& \theta(1 - p/\Lambda)~,\\ 
\label{LF} g_{{\rm L}\alpha}(p) &=& [1 + 
(p/\Lambda)^{2\alpha}]^{-1},\quad 
\alpha \geq 2~,\\ 
\label{GF} g_{\rm G}(p) &=& \exp(-p^2/\Lambda^2)~. 
\end{eqnarray} 
where $\Lambda$ is a cut-off parameter. These form factors are 
plotted in Fig.~\ref{FF_fig}. We achieve deviations from the NJL 
case (step function) by using the Lorentzian form factor with 
decreasing the $\alpha$ parameter. The Gaussian form factor 
appears on the other limit having a very soft momentum dependence. 
 
To perform numerical calculations one has to specify, for each 
form factor, the following set of parameters: the light quark {current} 
mass ($m$), the coupling strength ($G$) and the range of the 
interaction ($\Lambda$). The diquark coupling constant 
$H_v$ is fixed to the ratio $H_v/G=8/3$ in accordance with the result of 
the Fierz transformation for a one-gluon exchange interaction. 
In this work we use the parameterizations 
recently given in Ref.~\cite{Grigorian:2006qe} and listed in 
Table~\ref{parNJL}.
\begin{table}[htb] 
 \vspace*{0.5cm} 
\begin{center} 
\begin{tabular}{|c|c||c|c|c|}\hline 
$M$[MeV]  & Form factor &$\Lambda$[MeV] &$G~\Lambda^2$ &$m$[MeV]\\ 
\hline 
330          &NJL    &$629.5$ &$2.17$ &$5.28$\\ 
             &L10    &$649.2$ &$2.36$ &$4.71$\\ 
             &L5     &$666.5$ &$2.49$ &$4.09$\\ 
             &L3     &$685.8$ &$2.59$ &$3.25$\\ 
             &L2     &$703.4$ &$2.58$ &$2.37$\\ 
             &G      &$891.1$ &$3.88$ &$2.18$\\ 
\hline 
400          &NJL    &$587.9$ &$2.44$ &$5.58$\\ 
             &L10    &$600.3$ &$2.64$ &$5.01$\\ 
             &L5     &$609.3$ &$2.78$ &$4.39$\\ 
             &L3     &$616.2$ &$2.87$ &$3.55$\\ 
             &L2     &$617.8$ &$2.83$ &$2.65$\\ 
             &G      &$756.1$ &$4.22$ &$2.60$\\ 
\hline 
  \end{tabular} 
\caption{Parameter sets for the nonlocal chiral quark model with Lorentzian 
and Gaussian form factors and for the NJL model. Sets for different fixed $M=$ 
330, 400 MeV are listed.  } 
  \label{parNJL} 
\end{center} 
\end{table} 
 They have been obtained by fitting the vacuum properties of the 
pion ($f_{\pi }=92.4$~MeV, $M_{\pi }=135$ MeV) and the vacuum constituent 
quark mass at zero momentum, $M =m+\phi$. For the latter the 
phenomenologically reasonable values $M=$ 330 and 400 MeV are 
been used. 
 
Note that the results to be presented below 
do not depend on the choice of the Lorentzian-type function  as 
interpolating form factor. In fact, similar results have been 
obtained using other interpolating functions as, e.g., the 
Woods-Saxon form factors the parameterization of which is given 
in \cite{Grigorian:2006qe}.

\subsection{Quark mass and CSL pairing gap} 
 
\begin{figure}[bth] 
  \begin{center} 
   \includegraphics[width=1.\linewidth,height=0.4\textheight, 
angle = -90]{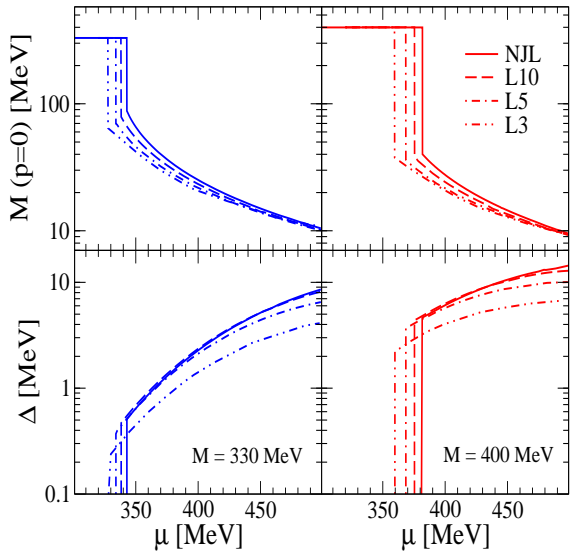} 
    \caption{The dependence of the dynamical mass $M(0)$ and the CSL pairing 
gap  $\Delta$ on the chemical potential $\mu$ for different NJL-like form factor 
models (NJL and Lorentzian with $\alpha \geq 3$). 
The parameterizations correspond to a fixed constituent mass: $M=330$ MeV 
on the left and $M=400$ MeV on the right panel.} 
    \label{FF_comparisn_1.3} 
  \end{center} 
\end{figure}

First, we analyze form factors which do not deviate strongly from the 
NJL case, i.e. L$\alpha$ for  $\alpha\geq 3$, shown as the grey area in 
Fig.~\ref{FF_fig}. In Fig.~\ref{FF_comparisn_1.3} 
we compare the solutions obtained for the mass and the CSL gaps for two 
different sets  of regularizations for fixed constituent  mass: 
$M=330$ MeV (left) and $M=400$ MeV (right). 
{For the parameterizations with a larger} constituent mass in 
vacuum, {one obtains a larger critical chemical potential $\mu_c$} 
for the phase transition from the chirally broken phase 
to the restored one, where the CSL pairing can occur. 
{The gaps at the onset, $\Delta(\mu_c)$, are larger 
whereas the mass gaps after the chiral transition are smaller.} 
 
Like in the NJL case \cite{Aguilera:2005tg}, the CSL gaps are strongly 
increasing functions of $\mu$ in the range that is relevant for compact 
stars, {$\mu_c < \mu \lsim 500$ MeV, where the upper limit is due 
to the threshold for the occurrence of strange quarks which allow 
pairing patterns like the CFL phase, more favorable than CSL} 
(for recent phase diagrams for neutral matter in the three flavor case see 
\cite{Blaschke:2005uj,Ruster:2005jc,Abuki:2005ms}). 
 
On the other hand, Fig.~\ref{FF_comparisn_1.3} clearly shows that 
the pairing gaps in this nonlocal extension are reduced relative to 
the NJL ones: the smoother the form factor, the smaller the gap. 
The reduction could be up to a factor three in the case of Lorentzian 
with $\alpha=3$.

But perhaps one of the most important effects of nonlocality 
from the point of view of the phenomenology of compact stars 
is the shift of the chiral phase transition to lower values of $\mu$. 
A lowering of the critical density for deconfinement makes 
stable hybrid star configurations with large quark matter cores possible 
\cite{Grigorian:2003vi,Shovkovy:2003ce}. 
On the other hand, results from lattice QCD simulations for the quark 
propagator \cite{Parappilly:2005ei} show a very  smooth four-momentum 
dependence of the quark self energies which is also a characteristics 
of confining quark models within the Dyson-Schwinger approach 
\cite{Blaschke:1997bj}. 
Smoother form factors could thus be more appropriate to model QCD interactions. 
{Note}, however, that the instantaneous 
nonlocal models {with such smooth} form factors lead to 
unrealistically large values of the chiral condensate (above $280$ 
MeV) in the vacuum \cite{Grigorian:2006qe}. 
 
\begin{figure}[bth] 
  \begin{center} 
\includegraphics[
width=1.\linewidth,height=0.4\textheight, angle = 
-90]{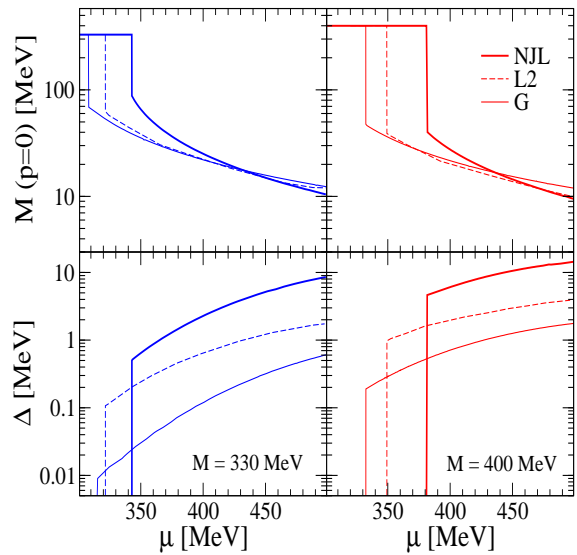} 
    \caption{Same as Fig.~2 for smooth form factor models 
(Lorentzian with $\alpha =2$ and Gaussian). 
} \label{FF_comparisn_1.4} 
  \end{center} 
\end{figure} 
 
Therefore we have subdivided our discussion of different form factors into 
two groups: those which lead to deviations from the NJL results within one 
order of magnitude and those resulting in larger deviations from 
the NJL model case (Lorentzian with $\alpha =2$ and Gaussian), see 
Fig.~\ref{FF_comparisn_1.4}. 
For example, the shift in the critical chemical potential for the onset 
of the chiral phase transition relative to the NJL case is less than 
$20$ MeV within the first group, but larger $30$ MeV for the second group, 
see Figs.~\ref{FF_comparisn_1.3} and \ref{FF_comparisn_1.4}. 
However, the qualitative behaviour of the chiral and CSL gaps is not affected 
by the choice of the form factors. 
 
\begin{figure}[bth] 
  \begin{center} 
   \includegraphics[width=1.\linewidth,height=0.4\textheight, 
angle = -90]{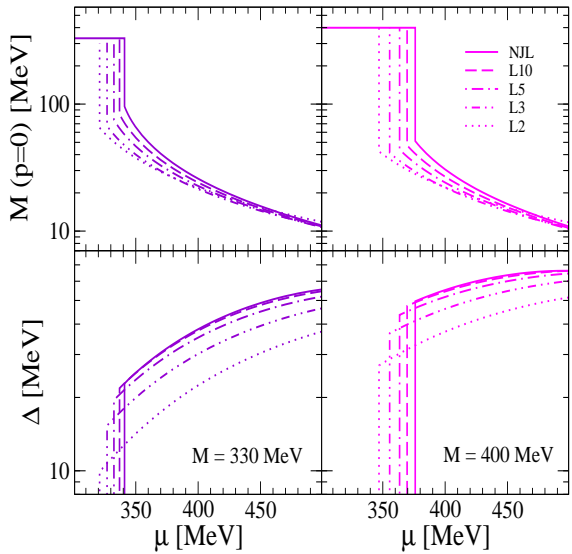} 
    \caption{Same as Fig.2 and Fig.3 but the coupling constant $H_v$ in the 
diquark channel is doubled with respect to the usual value coming from 
    Fierz transformed one gluon exchange interaction.} 
    \label{FF_comparisn_1.6} 
  \end{center} 
\end{figure}

To obtain the  above results, 
we have kept fixed the ratio $H_v/G$ 
 at the standard value obtained from Fierz transforming 
 the one-gluon exchange interaction. However, 
 in order to estimate the effect of possible 
 uncertainties in this value, we have considered 
 also the situation in which this ratio is taken 
 to be twice its Fierz value. The corresponding 
 results for $2H_v$ and all Lorentzian-type form factors 
 under consideration are shown in Fig.~\ref{FF_comparisn_1.6}. 
It is worth  noticing that while the increase of $H_v$ by a factor 
$2$ increases the CSL gaps by one ($M=400$ MeV) or two ($M=330$ MeV) orders of 
magnitude, 
the qualitative behaviour of $M$ and  $\Delta$ as functions of $\mu$ remains 
practically unchanged. 
The lowering of $\mu_c$ and $\Delta$ as the form factor becomes smoother 
is also similar to the previous case. 
 
\subsection{Quasiparticle excitation spectrum} 
\begin{figure}[bth] 
  \begin{center} 
\includegraphics[width=0.9\linewidth,height=0.4\textheight,angle = -90] 
{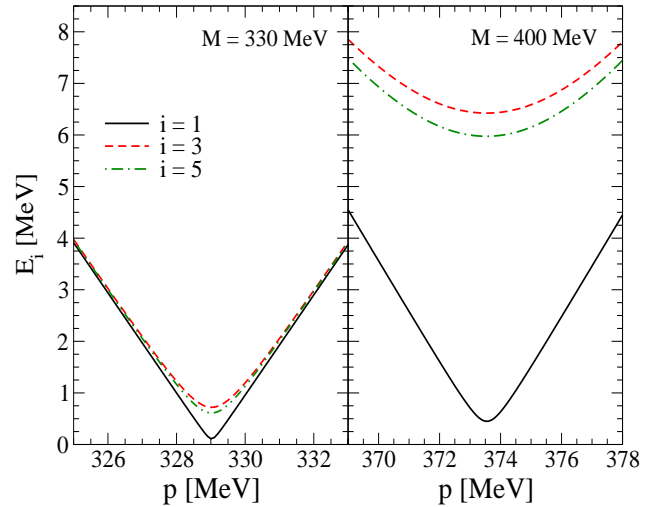} 
\caption{{Excitation energies in the CSL phase as a function 
    of the momentum 
for both parameterizations of the L10 model from Table I at the 
corresponding critical values of the chemical potential, $\mu = \mu_c$. 
Left panel: M=330 MeV, $\mu_c=338$ MeV and right panel: 
$M=400$ MeV, $\mu_c=375$ MeV. Antiparticle modes $E_2$, $E_4$ and 
$E_6$ are not shown.} 
} \label{Fig_dis} 
  \end{center} 
\end{figure}

In Fig.~\ref{Fig_dis} we show the quasiparticle excitation 
  spectrum at the critical chemical potential for both 
  parameterizations, respectively. 
{We observe that the quasiparticle mode with the lowest 
energy band corresponds is $E_{1}(p)$ with a minimum 
\bea 
E_{1,\rm min}= \min_p [E_1(p)]~, 
\label{Delta_minimum} 
\eea 
being the most relevant quantity for possible applications of the CSL phase of 
quark matter to compact star cooling phenomenology. This minimum 
occurs at the Fermi momentum $p=p_F$. In a very good approximation 
$p_F$ can be represented by the lowest orders of a series 
expansion in the parameter $s = p_F {g'_{L\alpha}(p_F)}/{g_{L\alpha}(p_F)}$, 
which is a measure for the influence of the form factor 
\begin{eqnarray} 
\label{pf} p_F^2 &=& \mu_{\rm eff}^2(p_F) - M_{\rm eff}^2(p_F)\nonumber\\ 
&& + 2 \Delta_{\rm eff}^2(p_F) 
\frac{M(p_F)[M(p_F)-m]+M_{\rm eff}^2(p_F)}{M^2(p_F)-M_{\rm eff}^4(p_F)/\mu^2} 
~s \nonumber\\ 
&& + O(s^2)~. 
\end{eqnarray} 
In the same order of the expansion in $s$, we obtain for the minimal excitation 
energy 
\begin{equation}\label{e1min} 
E_{1,\rm min} = \Delta_{\rm eff}(p_F)+ O(s^2)~. 
\end{equation} 
Although $E_{1,\rm min}$ might be quite small (see below) it never vanishes. 
In fact, as in the NJL case \cite{Aguilera:2005tg}, in the present class 
of models none of the dispersion relations lead to gapless modes. 
It is interesting to note that this lowest energy mode $E_1(p)$, relevant 
for compact star cooling phenomenology, is the one which is inert against 
the influence of a strong external magnetic field typical for neutron 
stars since it belongs to vanishing spin projection, $S_z=0$. 
  
In  Fig.~\ref{FF_effect} and Fig.~\ref{FF_effecta} we plot $E_{{1,\rm min}}$ 
as a function of $\mu$ for different models from the NJL-like and the 
smooth form factor groups, respectively. The calculations are made for 
both sets of parameterizations of Table 1, corresponding to constituent masses 
of $M=330$ MeV and $M=400$ MeV. 
According to the analytical approximative result of Eq. (\ref{e1min}), 
the behavior of the minimal excitation energies can be understood as a 
product of the increasing $\mu$-dependence of the CSL pairing gaps and 
the decreasing one of the other factors in Eq. (\ref{effective2}). 
For the parameterizations with $M=400$ MeV on the right panel we obtain 
that $E_{{1,\rm min}}$ is a decreasing function of $\mu$ 
since the increase in $\Delta(\mu)$ cannot 
  overcompensate the decrease in $M(p_F)g(p_F)/\mu_{\rm eff}(p_F)$. 
A similar effect has been reported for NJL models \cite{Aguilera:2005tg}. 
On the other hand, for parameterizations with $M=330$ MeV 
the interplay between $\Delta(\mu)$ and the other factors in 
Eq. (\ref{effective2}) is density dependent: 
a slightly increasing behavior of  $E_{{1,\rm min}}$ at 
low densities {is followed by} a tendency to a saturation or 
even decreasing behaviour at high densities. 
For the group of NJL-like form factors, our {results for} $E_{{1,\rm min}}$ 
lie in the range of $50-500$ keV and for the case of smooth form factors they 
are between $1$ and $100$ keV.

\begin{figure}[bth] 
  \begin{center} 
\includegraphics[width=0.9\linewidth,height=0.44\textheight,angle = -90] 
{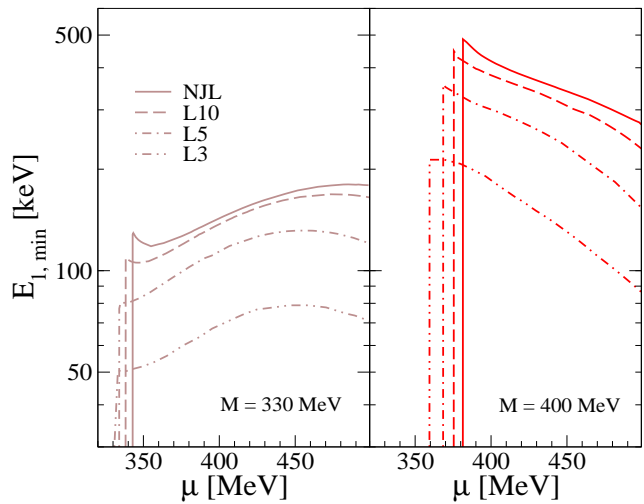} 
    \caption{Minimal excitation energies $E_{{1,\rm min}}$ in the CSL 
      phase as functions of 
    the chemical potential $\mu$  for different NJL-like form factors. 
Parameterizations correspond to fixed $M=330$ MeV on the left and 
$M=400$ MeV on the right. } 
\label{FF_effect} 
  \end{center} 
\end{figure} 
\begin{figure}[bth] 
  \begin{center} 
\includegraphics[width=0.9\linewidth,height=0.44\textheight,angle = -90] 
{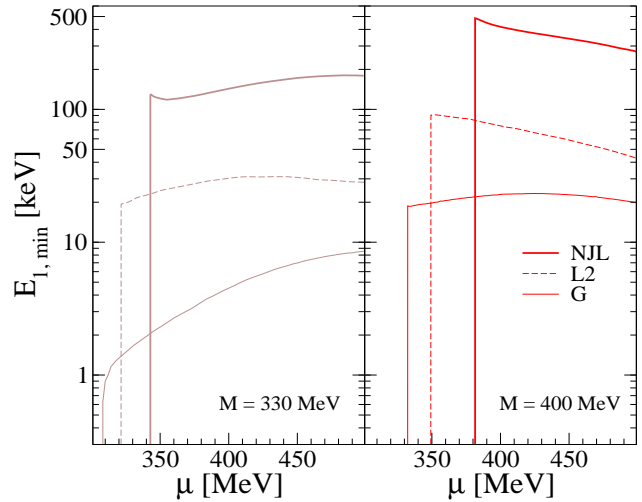} 
\caption{Same as Fig.~\ref{FF_effect} for models that present 
 a larger deviation from NJL models (Lorentzian with $\alpha =2$ and 
Gaussian). 
}\label{FF_effecta} 
  \end{center} 
\end{figure} 
  
Our main results are summarized in Fig.~\ref{Delta4}. 
In the upper panel 
we show the scaling of the 
the critical chemical potential $\mu_c^{\rm L\alpha}$ for the onset of the CSL phase 
for the Lorentztian-type model $\rm L\alpha$ with the 
smoothness parameter $1/\alpha$ 
normalized to the NJL limit case, $\mu_c^{\rm NJL}$. 
In the lower panel we show a comparative plot of 
the minimal excitation energies $E_{1,\rm min}^{\rm L \alpha}$ in units 
of the corresponding  NJL counterpart $E^{\rm NJL}_{{1,\rm min}}$, 
evaluated at $\mu_c^{\rm L\alpha}$ and $\mu_c^{\rm NJL}$, respectively. 
The corresponding 
results for $2H_v$ are also shown in Fig.~\ref{Delta4} as open symbols. As we see 
 the qualitative behaviour of both 
$\mu_c^{\rm L \alpha}/\mu_c^{\rm NJL}$ and
$E_{1,\rm min}^{\rm L \alpha}/E_{1,\rm min}^{\rm NJL}$ 
as a function of $1/\alpha$ 
 remains unchanged. 
 
It is remarkable that, as it would be expected from an expansion to 
lowest order in $s$, 
both quantities scale almost linearly with 
$1/\alpha$. In fact, 
a very good approximation to our numerical results is obtained with 
\bea 
\mu_c^{\rm L\alpha}
&\approx& \left(1-\frac{\xi^{\prime}}{\alpha}\right) \mu_c^{\rm NJL} ~, \\
E_{1,\rm min}^{\rm L\alpha}
&\approx& 
\left(1-\frac{\xi}{\alpha}\right)E^{\rm NJL}_{1,\rm min}  ~,
\eea 
for $\alpha$ down to 2, where the slope parameters $\xi$ and 
$\xi^{\prime}$  do only moderately depend on the model 
parameterization ($M$) and the CSL coupling strength  ($H_v$). 
For $M=330$ MeV we get $\xi$ = 1.8 (1.5) and $\xi'$ = 0.13 (0.12), 
while for $M=400$ MeV the corresponding values are $\xi$ = 1.6 (1.2), 
$\xi'$ = 0.17 (0.16).
The numbers in parentheses are obtained by doubling $H_v$.

\begin{figure}[htb] 
  \begin{center} 
\includegraphics[ 
width=1.\linewidth,height=0.4\textheight, 
angle = -90]{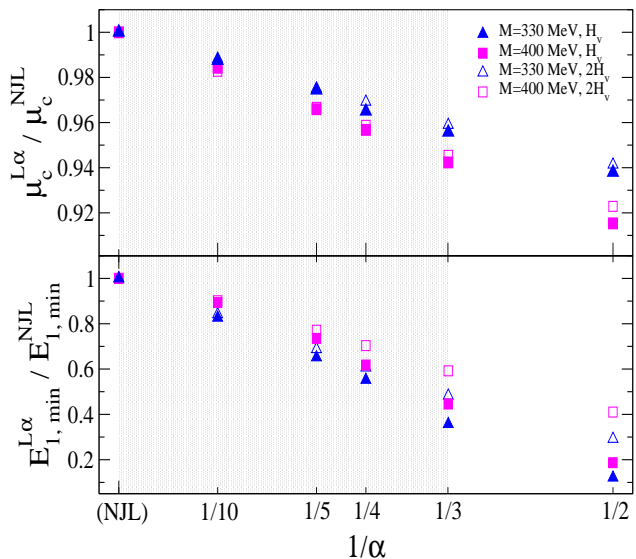} 
    \caption{Two  nonlocality effects on CSL condensates as a function 
of the smoothness parameter $1/\alpha$ of Lorentzian-type models, for $\alpha=2, 3, 4, 5, 10$,  normalized to the NJL limit ($\alpha \to \infty$).
First: the lowering  of the critical chemical potentials $\mu_c^{\rm L\alpha}$ 
respect to $\mu_c^{\rm NJL}$ (upper panel). 
Second: the reduction of the minimal excitation energies $E_{1,\rm{min}}^{\rm L\alpha}$ respect to  $E_{1,\rm{min}}^{\rm NJL}$
(upper panel; values calculated at 
$\mu_c^{\rm L\alpha}$ and  $\mu_c^{\rm NJL}$, respectively). 
Results are shown 
as a function of $1/\alpha$ for two different quark mass parameterizations 
in the vacuum: $M=330$ MeV (triangles) and $M=400$ MeV (squares).
The open symbols show results for doubled CSL coupling strength (2~$H_v$).} 
\label{Delta4} 
  \end{center} 
\end{figure} 

\section{Conclusion} 
 
 We have studied the effect of instantaneous nonlocal interactions in 
 the color spin locking (CSL) phase of quark matter. 
 We have introduced momentum dependent form factors to model the 
 nonlocality  and compared systematically with the local NJL counterpart. 
 
We have shown that there is a systematic lowering of the 
critical chemical potential for the onset of the CSL phase as well as for the 
minimal excitation energy (effective CSL gap) as a function of the nonlocality 
which can be represented as a linear dependence on the smoothness parameter 
$1/\alpha$ of the Lorentz-type form factor. 
These qualitative effects are shown 
to be robust under changes in the coupling constant 
used to represent the CSL interaction.
 
It has been found that hybrid star cooling requires 
all quark modes to be paired with a {minimal} pairing of 
the order of $~10-100$ keV to suppress the direct Urca process in quark 
matter. 
The present model for the CSL phase meets this requirement and calls for 
a more detailed analysis of the cooling phenomenology based on this 
microscopically justified pairing pattern. 
 
The smallest gap which governs the cooling phenomenology 
corresponds to the zero z-projection of the spin and thus remains 
unaffected by the external magnetic field of a compact star. 
Moreover, the CSL pairing pattern is a flavor singlet and 
insensitive to the flavor asymmetry in a compact star under 
$\beta$-equilibrium.
 
Therefore, the CSL phase with nonlocal instantaneous interactions is 
particularly interesting for applications in compact stars and allows 
to achieve a suitable description of 
quark matter properties by choosing the appropriate form factor models. 
Although it remains to be shown that hybrid star configurations with 
the CSL quark matter phase could be stable, due to the small gaps, we 
expect to have results reproducing those of the normal quark matter 
case, where stable quark matter cores in nonlocal models have been 
found.

\section*{Acknowledgments} 
 
We thank J. Berdermann, M. Buballa, T. Kl\"ahn, D. Rischke, 
D.N. Voskresenksky and Q. Wang for their discussions and interest in our 
studies. 
D.N.A. acknowledges interesting discussions  with J.~Pons on this work. 
D.N.A. is grateful for the hospitality at the GSI Darmstadt and to the 
University of Rostock where this work was started and most of the calculations have been performed. 
The work was supported by the  Virtual Institute 'Dense hadronic matter and 
QCD phase transitions' of the Helmholtz Association under grant 
VH-VI-041, and  by a scientist exchange program 
between Germany and Argentina funded jointly by DAAD under 
grant No.\ DE/04/27956 and ANTORCHAS under grant No.\ 4248-6. 
D.N.A received support from VESF Fellowships EGO-DIR-112/2005, 
N.N.S. has been supported in part by CONICET and ANPCyT 
(Argentina), under grants PIP 6084 and PICT00-03-08580. 
The work of H.G. has been supported by DFG under grant number 436 ARM 
17/4/05.

\end{document}